\documentclass[aps,twocolumn,showpacs,showkeys,floatfix,longbibliography,superscriptaddress]{revtex4-1}

\usepackage{graphicx}
\usepackage{textcomp}
\usepackage{dcolumn}
\usepackage{amsmath}
\usepackage{amssymb}
\usepackage{epstopdf}
\usepackage{rotating}
\usepackage{color}
\usepackage{natbib}
\usepackage{url}
\setcitestyle{square}
\usepackage{textgreek}
\usepackage{multirow}
\usepackage{makecell}
\usepackage{setspace}
\usepackage{etoolbox}
\usepackage{lipsum}
\usepackage{booktabs}
\usepackage{array}
\usepackage{hyperref}  
\usepackage[hypcap=true]{caption}
\hypersetup{colorlinks=true,urlcolor=blue,citecolor=blue,linkcolor=blue}   
\usepackage{enumitem}          
\setlist{nosep}                 

\usepackage{enumitem}

\begin{document}

\title{Online Administration of Research-Based Assessments}


\author{Ben Van Dusen}
\affiliation{School of Education, Iowa State University, Ames, IA, 50011, USA}

\author{Mollee Shultz}
\affiliation{Department of Physics, Texas State University, San Marcos, TX, 78666, USA}

\author{Jayson M. Nissen}\affiliation{Nissen Education Research and Design, Corvallis, Oregon, 97333, USA} 

\author{Bethany R. Wilcox}
\affiliation{Department of Physics, University of Colorado Boulder, Boulder, CO 80309, USA}

\author{N.G.~Holmes}
\affiliation{Laboratory of Atomic and Solid State Physics, Cornell University, Ithaca, NY 14850, USA}

\author{Manher Jariwala}
\affiliation{Department of Physics, Boston University, Boston, MA 02215, USA}

\author{Eleanor W. Close}
\affiliation{Department of Physics, Texas State University, San Marcos, TX, 78666, USA}

\author{Steven Pollock}
\affiliation{Department of Physics, University of Colorado Boulder, Boulder, CO 80309, USA}

\begin{abstract}
The number and use of research-based assessments (RBAs) has grown significantly over the last several decades. Data from RBAs can be compared against national datasets to provide instructors with empirical evidence on the efficacy of their teaching practices. Many physics instructors, however, opt not to use RBAs due to barriers such as having to use class time to administer them. In this article we examine how these barriers can be mitigated through online administrations of RBAs, particularly through the use of free online RBA platforms that automate administering, scoring, and analyzing RBAs (e.g., the Learning About STEM Student Outcomes [LASSO] \citep{LASSO}, Colorado Learning Attitudes About Science Survey for Experimental Physics [E-CLASS] \citep{ECLASS}, Physics Lab Inventory of Critical thinking [PLIC] \citep{PLIC}, and PhysPort DataExplorer \citep{dataexplorer} platforms). We also explore the research into common concerns of administering RBAs online and conclude with a practical how-to guide for instructors. 
\end{abstract}
\maketitle

\section{Introduction}
Research-based assessments (RBAs) are instruments designed to measure the impact of a course on student outcomes, such as content knowledge, attitudes, and identities. Physics instructors and researchers have used RBAs (e.g., the Force Concept Inventory~\citep{Hestenes1992}) to examine the impact of courses and to inform research-based pedagogical practices \citep{mazur2013peer, chabay2015matter, madsen2019resource, madsen2017resource, von2016secondary, crouch2001peer, wilcox2014coupled, van2020equity, redish2004teaching, meltzer2012resource, national2012discipline, hake1998interactive, ding2013students, crouch2001peer,chasteen2015sei}. Unlike other course exams, the common usage of standardized RBAs across institutions uniquely supports instructors to compare their student outcomes over time or against multi-institutional datasets and also supports large-scale PER investigations.

Using RBAs began as an activity in which only a few instructors engaged, particularly those interested in PER. Instructors face several logistical barriers to administering RBAs but, as the number and use of RBAs have expanded, so have resources that support instructors in using them. In particular, online platforms are now available to help instructors throughout the process from selecting (e.g., PhysPort \citep{PhysPort}), administering, scoring, and analyzing RBAs (e.g., the Learning About STEM Student Outcomes [LASSO] \citep{LASSO}, Colorado Learning Attitudes About Science Survey for Experimental Physics [E-CLASS] \citep{ECLASS}, and Physics Lab Inventory of Critical thinking [PLIC] \citep{PLIC} platforms). These online platforms remove or lower the barriers to using RBAs, making it easy for even first-time users to measure their students' outcomes systematically. As more courses transition to be offered online, these web-based systems are increasingly useful.

We hope that this paper can serve as a guide for instructors considering administering RBAs online. In what follows, we examine common barriers to using RBAs, how online administration can ameliorate those barriers, and the research into online administration of RBAs. We also include a practical how-to for administering RBAs online, and, in the appendix, we include sample email wording for pretest and posttest administrations.

\section{Barriers to using RBAs on paper, and online solutions}
Below we have listed common reasons instructors give for choosing {\it not} to use RBAs during class and explanations for how online administration address these concerns.
\par \textbf{I can't spare 30+ minutes of class time twice in a semester to give an RBA.} Administering RBAs online allows students to complete RBAs either at home or in class. Studies have found that with sufficient incentives, students' participation and scores are the same whether completed in class or at home (see the discussion in the next section).
\par \textbf{I don't have the time or TA power to score an RBA.} Administering the RBA online removes the step of scoring scantrons or paper surveys and automatically generates spreadsheets of student responses that can be quickly and easily analyzed. Online RBA platforms (e.g., LASSO \citep{LASSO}, E-CLASS \citep{ECLASS}, PLIC \citep{PLIC}, and PhysPort DataExplorer \citep{dataexplorer}) can automate the scoring process altogether, providing instructors full student responses and scored responses. 
\par \textbf{I need an online version of the assessment and can't spare the time to set this up myself.} Online RBA platforms already host and administer a wide array of physics RBAs for free.
\par \textbf{I don't know what my results mean.} Online RBA platforms can automatically generate reports that include visualizations and summary statistics that contextualize student outcome data.  This can help instructors make sense of their students' performance and inform concrete changes to their instruction.
\par \textbf{I don't have access to any comparison data.} Online RBA platforms can standardize data formats, making it easy to compare or combine course data. These platforms collect course meta-data that can also be used to identify appropriate comparison points for a wide range of courses and institutions.  They can also automatically aggregate and anonymize datasets to support PER in performing large-scale, multi-institution investigations. 

\section{Barriers to using RBAs online, and research-based responses}

Moving an RBA online and at-home, particularly one targeting content knowledge as opposed to attitudes and beliefs, brings with it several potential concerns, particularly around issues of student engagement, test security, and use of unauthorized resources. Below, we articulate some of these concerns and summarize research findings that begin to address them. Note that the findings discussed here represent a snapshot of the current state of understanding about student engagement with online RBAs; as the use of online tests becomes more common and norms change, these findings may become less generalizable. 

\textbf{Does giving the test online impact how many and which of my students participate?}
Low-stakes RBAs administered online have yielded similar participation rates as the equivalent paper tests administered in class~\cite{nissen2018participation}. In an experiment where researchers randomly assigned students (\textit{N}=1310) at one institution to take the same RBA online outside of class versus on paper inside class, participation rates were comparable if instructors administered the RBAs using the recommended practices described in Sec.~\ref{sec:howto} (also accessible at ~\cite{PhysPortRecommend}). Moreover, the participation rates did not differ between online and in-person based on gender or final course grade \cite{nissen2018participation}. Incentive structures strongly influence participation rates; for example, another study \cite{wilcox2019online} found an increase in online participation rates compared to historical norms, attributed to changes in incentives (explicit credit for participation when administered online). 

\textbf{Does taking the test online impact the score for my course and can I compare my scores from online versions to my scores from in-person versions or published results?}
In the first study described above~\cite{nissen2018participation}, researchers found that student performance on the online, computer-based tests were equivalent to performance on the same tests administered on paper during class~\cite{nissen2018participation}. This result held for both concept inventory tests and attitudinal surveys, suggesting that instructors can compare results from online and in-person administrations. The second study described above~\cite{wilcox2019online} found slightly lower online scores relative to historical data sets. They attributed this effect to the increased participation rate from lower-performing students in online assessments compared with in-person assessments.  This result suggests a reduction in the common sampling bias toward higher-performing students and would make the scores more representative. 

\textbf{What if my students use the internet to look up the answers to the questions?}
In a study examining students’ behaviors when taking research-based assessments online \cite{wilcox2019online}, researchers found that only ${\sim}10\%$ of students showed direct evidence of copying question text, potentially intending to search the text to find the correct answer online. For tests with solutions readily available online, this behavior correlated with increased performance, while for tests without available solutions, it correlated with lower performance. However, because the proportion of students engaging in these behaviors was small, the impact on the overall average for the course was not significant. These findings align with other~\cite{nissen2018participation} findings about the lack of impact on performance associated with administering an assessment online.

\textbf{What if my students get distracted and don’t take the test seriously?}
Researchers have used browser focus data (i.e., how often and for how long the assessment tab becomes hidden on the student’s screen) to determine how common distraction might be during online RBAs. This study~\cite{wilcox2019online} found that browser focus data indicated that between half and two-thirds of students lost focus on the assessment at least once, though the majority of these events (two-thirds) was less than 1 min in duration. Additionally, neither the number nor the duration of focus loss events correlated with students’ scores. Thus, in that study, there was no apparent negative impact on students’ scores due to distraction in the online environment. 

\textbf{What if my students save the test and post it online?}
Security of research-based assessment is an issue that becomes particularly important when administering the assessments online, and, in practice, the nature of these concerns depends on the assessment in question. For example, well-used introductory assessments such as the FMCE or BEMA are already available online on paid sites such as Chegg or CourseHero \cite{wilcox2019online}. Less well used or newer assessments do not appear to have worked solutions available online to date. In one study, very few students (less than 1-2\%) attempted to save the test using print commands during online assessments~\cite{wilcox2019online}. However, it is likely inevitable that questions (and solutions) will become increasingly available to students over time. This makes it all the more important that faculty keep these assessments low-stakes, not graded, and provide appropriate instructions to motivate students to take the assessment in the intended spirit, as a learning tool (see the next section).

\section{Practical how-to}\label{sec:howto}

Many of the strategies for implementing RBAs on paper~\cite{Madsen2017TPT} also apply to implementing RBAs online, though there are some unique strategies.

\par \textbf{Provide points-based incentive:} For points-based incentives, the point value should be small enough to keep participation low-stakes, but large enough to meaningfully motivate participation. We recommend on the order of a few percent of the total grade or the equivalent of a small portion of a homework assignment. Only give points for participation, not for correctness. Be sure to emphasize this to students through your class communications and in your syllabus.

 
\par \textbf{Provide multiple reminders \footnote{We have included a sample script for an in-class or email announcement in the appendix}:} If students are completing the surveys in their own time (not during class), we recommend sending multiple email reminders leading up to the deadline and making multiple announcements during class time. The repeat announcements help catch students who may have missed the first notification and indicate to students that you, as an instructor, value their participation. Because participation is often higher on pretests than posttests~\cite{nissen2018participation}, we suggest sending more notifications and re-emphasizing incentives at posttest.

\par \textbf{Use dedicated class time:} Some instructors may be uncomfortable providing points-based incentives, in which case it may be easier to use class time. Research has indicated that participation rates are similar whether the students complete the RBAs during or outside of class time (with appropriate incentives). While unnecessary, it is still appropriate to use time during a scheduled class session for students to complete the surveys online. To administer instruments online, instructors can share the link during the class session (such as during the first lab or tutorial).

\par \textbf{Communicate the goals:} When announcing the RBA to students, explicitly explain the goal is to obtain important information about the course and the instruction to better serve the students in the class (in the present and the future). Briefly describe the benefit of their participation to you, the instructor (in terms of feedback for the course), and to them, the students (in terms of study opportunities). Explicitly state that the goal is not to evaluate the students individually. Encourage students to answer all the questions, even if they are not confident in their responses, so you can adapt instruction accordingly. 

\par \textbf{Refer to the surveys through generic names:} When describing the RBA, use a generic name (such as ``course survey'') rather than the official instrument name so students can less easily search for the instrument online. 

\par \textbf{Avoid enforcing time limits:} Although many RBAs have recommended time limits, placing time limits on the instruments themselves can increase students' test anxiety and sense of higher stakes. Not placing strict time limits can also mediate technical difficulties students may face with online administration.

\par \textbf{Do not share the solutions, answers, or students' scores:} This helps maintain the security of the RBA so the community can continue to use it~\cite{redish2004teaching}. Providing scores can motivate students to want to know the solutions to the problems. The scores themselves are likely difficult for students to interpret. For example, pre-scores are typically quite low (in some cases post-scores as well), which could demoralize students without the appropriate context. Furthermore, RBA scores are only informative about group-level scores, not individual students, making individual scores less useful. 

\section{Conclusions}
RBAs are useful measures of the impact of a course on students' conceptual learning or attitudes/beliefs and have been a major driver of change in physics education. Many physics instructors, however, do not use RBAs for a variety of reasons. Online administration of RBAs can remove many of the barriers to administering, scoring, and analyzing RBA results. Researchers have found that instructors can get similar amounts and quality of RBA data whether they administer them in-class or online. Further, research to date has found minimal impact in student scores from using unauthorized resources or evidence of students compromising assessment security when administered online. Instructors can administer RBAs online by setting up their own online version (e.g., in their learning management system) or they can use existing online RBA administration platforms (e.g., LASSO \citep{LASSO}, E-CLASS \citep{ECLASS}, and PLIC \citep{PLIC}). These platforms 
are free and offer several advantages, such as lowering setup time, automatically generating reports, and contributing to large-scale PER investigations.

\section{References}
\bibliography{Bibliography.bib}

\section{Appendix: Sample script}
Suggested script for an in-class or email announcement:

\textbf{Pre-test:} You will receive emails from me with links to different pre-test surveys (a “concepts” survey and a “nature of physics” survey), which are part of your first homework assignment. We ask you to answer these pre-test questions to give us a better idea of your understanding of physics before the class begins. We use your responses to the survey to tailor our instruction in the course - not to evaluate you. If you are unsure about your answers, do not worry - this is useful information for us. Please try your best to answer the questions without help from any textbooks or anyone else. You will receive full credit (equal to one-half of a homework assignment) just for answering all the questions, right or wrong. Please complete the two pre-tests by ...
\par \textbf{Post-test:} You will receive emails from me with links to different post-test surveys (a “concepts” survey and a “nature of physics” survey), which are part of your last homework assignment. As with the surveys at the start of the course, we use your responses to the post-survey to evaluate our instruction in the course - not to evaluate you. If you are unsure about your answers, do not worry - this is useful information for us. Please try your best to answer the questions without help from any textbooks or anyone else and use the survey as an opportunity to test your own understanding to help guide your studying for the final exam. You will receive full credit (equal to one-half of a homework assignment) just for answering all the questions, right or wrong. Please complete the two post-tests by ...
\end{document}